\documentclass[11pt,a4paper]{article}

\usepackage{amsmath,amssymb}
\usepackage{graphicx}
\usepackage{authblk}
\usepackage[numbers]{natbib}
\usepackage{geometry}
\usepackage{float}
\usepackage[colorlinks=true, 
            linkcolor=blue, 
            citecolor=blue, 
            urlcolor=blue]{hyperref}

\title{\textbf{Testing the Constancy of Type Ia Supernova Luminosities with Gaussian Process}}

\author[1]{Akshay Rana\footnote{Email: akshay@ststephens.edu}}
\affil[1]{St. Stephen's College, University of Delhi, Delhi 110007, India}
\date{\today}

\begin{document}

\maketitle


\begin{abstract}
Type Ia supernovae (SNe~Ia) are central to studies of cosmic expansion, under the assumption that their absolute magnitude $M_B$ does not evolve with redshift. Even small drifts in brightness can bias cosmological parameters such as $H_0$ and $w$. Here we test this assumption using a non-parametric Gaussian Process (GP) reconstruction of the expansion history from cosmic chronometer $H(z)$ data, which provides a model-independent baseline distance modulus, $\mu_{\rm GP}(z)$. To propagate uncertainties, we draw Monte Carlo realizations of $H(z)$ from the GP posterior and evaluate them on a Chebyshev grid, which improves numerical stability and quadrature accuracy. Supernova observations are then compared to this baseline through residuals, $\Delta M_B(z)$, and their derivatives. Applying this method to Pantheon+ (1701 SNe~Ia) and DES 5YR (435 SNe~Ia), we find that SNe~Ia are consistent with being standard candles within $1\sigma$, though both datasets exhibit localized departures: near $z \sim 1$ in Pantheon+ and at $z \sim 0.3$--$0.5$ in DES. The presence of similar features in two independent surveys suggests they are not purely statistical. Our results point toward a possible non-monotonic luminosity evolution, likely reflecting different physical drivers at different epochs, and highlight the need for a deeper astrophysical understanding of SN~Ia populations.
\end{abstract}

\section{Introduction}

Type Ia supernovae (SNe Ia) are the cornerstone of modern observational cosmology. The discovery of cosmic acceleration through SN Ia measurements \citep{Riess_1998, 1999parl} established them as powerful distance indicators and placed dark energy at the center of cosmological research. In the decades since, steadily growing and increasingly homogeneous SN Ia compilations have extended the Hubble diagram to redshifts $z > 2$, providing some of the most stringent constraints on the cosmic expansion history and the dark energy equation of state \citep{2013wein, Scolnic_2018, scolnic2022}. The entire enterprise rests on the assumption that SNe Ia, after standardization, share a redshift-independent absolute magnitude, $M_B$ \citep{branch1993,phillips1993}. 

However, the constancy of $M_B$ is not guaranteed. A variety of astrophysical factors can, in principle, imprint a redshift dependence: the relative contributions of single- and double-degenerate progenitor channels \citep{Hayden_2010,Ruiter_2024}, correlations with host galaxy properties such as stellar mass and star-formation rate \citep{Gar_2005, Rose2020}, the age and metallicity evolution of progenitor populations \citep{Lee_2021, How_2009,Man_2006}, and potential effects of intergalactic dust or exotic opacity sources \citep{Nair_2012, Holanda_2018, Goobar_2018}. Even a mild evolution at the $\sim 0.01$--$0.05$ mag level can bias cosmological inference, shifting $w$ and other parameters at the significant level \citep{Campbell_2016, Linden_2009, sapone2021}. The question of whether SNe Ia remain true ``standard candles'' across cosmic time is therefore both astrophysically and cosmologically significant. 

Numerous studies have sought to test for possible luminosity evolution in Type~Ia supernovae. A common strategy has been to impose simple parametric forms - linear, quadratic, logarithmic, exponential, or power-law dependencies and to constrain their amplitudes with current data \citep{Seikel2008,tutusaus2017,darshan2022}. These analyses typically conclude that there is no statistically compelling evidence for strong monotonic evolution within present uncertainties. Yet, because these parametrizations enforce overly simple functional forms, they can overlook faint, scale-dependent, or non-monotonic effects arising from progenitor demographics or environmental conditions. To address this, alternative methodologies have been developed, including non-parametric reconstructions of the Hubble diagram and Bayesian hierarchical models of SN populations \citep{Rubin_2025,Gray_2024,wojtak_2023,mohlabeng_2014}. Others have pursued cross-checks with independent cosmological probes, explicitly testing the standard-candle assumption in a joint framework \citep{Linden_2009,Dinda_2023}. Collectively, these works reinforce the overall robustness of SN cosmology, but they also underscore a key limitation: the possibility remains that subtle, redshift-dependent variations are being smoothed out or misinterpreted under overly simplistic modeling assumptions. It is precisely this open space between the null evidence for strong evolution and the lingering degeneracies with cosmology that motivates the present work.

In this work we revisit the question of SN Ia luminosity evolution with a complementary, model-independent strategy. Rather than assuming a cosmological model to predict distances, we reconstruct the expansion history directly from cosmic chronometer measurements of $H(z)$ \citep{jimenez_2002, Moresco_2012, Moresco_2016,rana_2017}, and from this obtain the corresponding distance modulus, $\mu_{\rm GP}(z)$, via Gaussian Processes. The deviation between observed supernova moduli and this reconstruction, \(\Delta M_B(z) \equiv \mu_{\rm obs}(z) - \mu_{\rm GP}(z)\), then serves as a direct probe of luminosity evolution. By additionally examining the derivative $d\Delta M_B/dz$, we are able to trace not only the presence but also the \emph{rate of change} of possible evolution, enhancing sensitivity to subtle signatures. 

Our analysis draws primarily on the Pantheon+ compilation \citep{scolnic2022}, which provides the largest and most homogeneous set of 1701 spectroscopically confirmed SNe Ia to date. To substantiate and cross-check our findings in the overlapping redshift range, we further employ the Dark Energy Survey five-year supernova sample (DES-SN5YR), consisting of 435 spectroscopically confirmed SNe Ia \citep{DES_2024,Camilleri_2024}. While smaller, the DES dataset offers an independent calibration and homogeneous survey design, making it an important complementary check on Pantheon+ results\citep{Vincenzi_2025}. 

The novelty of this work lies in three aspects. 
\begin{itemize}
    \item First, we define the deviation $\Delta M_B(z) \equiv \mu_{\rm obs}(z) - \mu_{\rm GP}(z)$ as a direct diagnostic of possible SN Ia luminosity evolution, and extend the analysis to its derivative $d\Delta M_B/dz$, thereby probing not only the amplitude of possible deviations but also their redshift dependence, providing a sharper and more sensitive test of evolutionary signatures.
    \item Second, we reconstruct the Hubble parameter $H(z)$ from cosmic chronometer measurements using Gaussian Processes, and obtain the corresponding luminosity distance by propagating the full covariance through 2000 Monte Carlo realizations, thereby capturing both statistical noise and correlated systematics. For numerical stability and to reduce integration errors, the GP posterior is evaluated on Chebyshev nodes, which minimizes interpolation artifacts and yields a stable, smooth representation of $\mu_{\rm GP}(z)$ for comparison with supernova data. This framework ensures that uncertainties are propagated consistently from $H(z)$ measurements to the derived distance modulus in a fully non-parametric and model-independent manner.

    \item Third, we perform this test on two independent supernova samples: the Pantheon+ compilation as our primary dataset, and the spectroscopically confirmed DES-SN5YR sample as a homogeneous and independent cross-check. The joint use of these datasets enables us to substantiate any detected trends and assess their robustness across different survey strategies. 
\end{itemize}

Taken together, these elements provide a methodology cum cosmology balance: the methodological advance of an MCMC-based GP integration scheme, and the cosmological insight gained from contrasting multiple SN datasets to search for fine structure in $\Delta M_B(z)$. This dual perspective allows us to move beyond parametric fits and test the standard candle assumption of SNe Ia with unprecedented sensitivity. In Section~\ref{method}, we describe the methodology, including the mathematical framework and the statistical technique of Gaussian Processes. Section~\ref{Data} provides details of the datasets employed, along with their underlying assumptions. The results, together with the corresponding plots, are presented in Section~\ref{Result}, followed by a discussion in Section~\ref{Discussion}.

\section{Methodology}\label{method}

\subsection{Distance Modulus as a Diagnostic}

In supernova cosmology, the distance modulus provides the fundamental connection between observations and theory.  The observable quantity entering cosmological analyses  is

\begin{equation}
\mu_{\rm obs}(z) = m_B(z) - M_B(z).
\end{equation}

Here, $m_B(z)$ corresponds to the  rest-frame peak $B$-band magnitude of SNe Ia,  and the absolute magnitude is denoted by$M_B(z)$.

From the theoretical side, the corresponding prediction is
\begin{equation}
\mu_{\rm th}(z) = 5\log_{10}\!\left(\frac{d_L(z)}{\mathrm{Mpc}}\right) + 25,
\end{equation}
with the luminosity distance in flat space time
\begin{equation}
d_L(z) = (1+z)\,r(z), \qquad 
r(z) = \int_0^z \frac{dz'}{H(z')}.
\end{equation}
Thus, every cosmological inference with supernovae reduces to the  comparison between $\mu_{\rm obs}(z)$ and $\mu_{\rm th}(z)$.  Once $\mu_{\text{th}}(z)$ is specified, we define the residuals as

\begin{equation}
\Delta \mu(z) \equiv \mu_{\text{obs}}(z) - \mu_{\text{th}}(z).
\end{equation}

This residual directly encodes any mismatch between the observed  brightness of SNe Ia and the theoretical prediction.  If SNe Ia are truly standard candles with a redshift-independent  absolute magnitude $M_B = M_{B0}$, then $\Delta\mu(z)$ should be  consistent with zero apart from statistical scatter and measurement noise.  A different choice of the fiducial $M_{B0}$ would simply shift all  $\Delta\mu(z)$ values by a constant offset, leaving their redshift  dependence unchanged. Thus, only systematic departures of $\Delta\mu(z)$  with redshift carry physical significance: they would indicate evolution  in the intrinsic luminosity of SNe Ia or unaccounted astrophysical effects,  rather than a calibration uncertainty in the absolute magnitude zero-point.

To make this explicit, let us parametrize the absolute magnitude as
\begin{equation}
M_B(z) = M_{B0} + \Delta M_B(z),
\end{equation}
where $M_{B0}$ is the fiducial constant absolute magnitude adopted in 
supernova cosmology through light-curve standardization, 
and $\Delta M_B(z)$ represents any redshift-dependent deviation from this value. 
Substituting this expression into the definition of $\mu_{\rm obs}(z)$ gives
\begin{align}
\mu_{\rm obs}(z) &= m_B(z) - \big(M_{B0} + \Delta M_B(z)\big) \\
                 &= \underbrace{\big[m_B(z) - M_{B0}\big]}_{\mu_{\rm th}(z)} - \Delta M_B(z).
\end{align}
Taking the difference with $\mu_{\rm th}(z)$, the residual becomes
\begin{equation}
\Delta\mu(z) \equiv \mu_{\rm obs}(z) - \mu_{\rm th}(z) = - \Delta M_B(z).
\end{equation}

This relation has a simple but powerful implication: any systematic structure in the residuals is equivalent to an evolving absolute magnitude. A flat residual function indicates that the standard candle assumption holds, while a redshift-dependent trend would signal evolution in SN Ia luminosity.  The minus sign has a straightforward interpretation: if supernovae at higher redshift are intrinsically brighter than the fiducial value ($\Delta M_B(z)<0$), then the observed modulus appears larger and $\Delta\mu(z)>0$. Conversely, if they are dimmer ($\Delta M_B(z)>0$), the residuals become negative. Thus, the shape of $\Delta\mu(z)$ directly encodes the presence and direction of possible luminosity evolution.

In standard analyses, $\mu_{\rm th}(z)$ is obtained by assuming a fiducial cosmological model, most often flat $\Lambda$CDM. This provides a useful baseline, but it also introduces a degeneracy: residuals relative to $\mu_{\rm obs}(z)$ may arise either from genuine supernova luminosity evolution or from the assumptions built into the cosmological model itself. To avoid this model dependence, we replace $\mu_{\rm th}(z)$ with a non-parametric reconstruction $\mu_{\rm GP}(z)$ derived directly from cosmic chronometer measurements of the Hubble parameter $H(z)$. The $H(z)$ data are smoothed using Gaussian Process regression, and the comoving distance integral is evaluated numerically, thereby propagating uncertainties consistently. The details of this procedure are presented in the following section. For the present discussion, we emphasize the resulting diagnostic,
\begin{equation}
\Delta M_B(z) \equiv \mu_{\rm obs}(z) - \mu_{\rm GP}(z),
\end{equation}
which is anchored solely in the empirical expansion history, free from cosmological priors.

Further, the observed distance modulus $\mu_{\rm obs}(z)$ is obtained through light-curve standardization, most commonly using the SALT2 model. The rest-frame peak $B$-band magnitude $m_B$ is corrected for light-curve shape ($X_1$), color ($c$), and host-galaxy mass dependence ($\Delta_M$), leading to the standardized form
\begin{equation}
\mu_{\rm obs}(z) = m_B - M_{B0} + \alpha X_1 - \beta c + \Delta_M,
\end{equation}

where $\alpha$ and $\beta$ are global nuisance parameters calibrated by the survey, and $\Delta_M$ accounts for the empirical ``host-mass step'' correction. This procedure ensures that SNe~Ia form a homogeneous sample for cosmology, but it also introduces implicit assumptions: the parameters $\alpha$, $\beta$, and $\Delta_M$ are usually treated as redshift-independent. If these parameters themselves evolve, the effect can masquerade as a signal of luminosity evolution.

Hence, the residual function $\Delta M_B(z)$, together with its derivative 
$d\Delta M_B/dz$, forms a two-tiered diagnostic of the standard candle hypothesis:
\begin{itemize}
    \item A flat $\Delta M_B(z)$ across redshift is consistent with constant $M_B$, reaffirming the validity of SNe Ia as standardizable candles. 
    \item A monotonic drift in $\Delta M_B(z)$, or a persistent  nonzero bias in $d\Delta M_B/dz$, would suggest gradual astrophysical evolution such as progenitor aging or metallicity enrichment. 
    \item Localized deviations or slope changes could signal more complex  systematics, e.g.\ varying progenitor channels or correlations with host galaxy environments. 
    \item A positive slope $\big(d(\Delta\mu)/dz > 0\big)$ corresponds to a decrease in the absolute magnitude with redshift $\big(dM_B/dz < 0\big)$, meaning that supernovae appear progressively \textit{brighter} at given redshifts. Conversely, a negative slope $\big(d(\Delta\mu)/dz < 0\big)$ implies an increase in the absolute magnitude with redshift $\big(dM_B/dz > 0\big)$, indicating that supernovae become \textit{fainter} at that redshifts.

    \item Sharp features in $d\Delta M_B/dz$ may reveal narrow redshift windows where distinct physical effects dominate, such as dust opacity, exotic photon interactions, or transitions in stellar populations. 
\end{itemize}

This framework elevates residual analysis beyond a simple offset test. By examining both the cumulative deviations and their differential behavior, we obtain a sharper and model-independent lens on whether subtle astrophysical or exotic effects imprint themselves on the luminosities of Type Ia supernovae.

\subsection{Non-Parametric Reconstruction with Gaussian Processes}

Parametric approaches to probing the redshift evolution of the absolute magnitude of Type Ia supernovae, such as linear or logarithmic models, impose specific functional forms on $\Delta M_B(z)$. While straightforward, such assumptions risk either overlooking subtle signatures or introducing artificial structure not supported by the data. To avoid these limitations, we adopt a \emph{non-parametric} framework based on Gaussian Process (GP) regression, a method widely applied in cosmology to reconstruct functions such as the Hubble parameter $H(z)$, the dark energy equation of state, and distance duality relations \citep{seikel2012, shafieloo2012,Hwang_2023}. In this work, GPs are employed to reconstruct the cosmic expansion history $H(z)$ from cosmic chronometer measurements, to propagate uncertainties consistently through integrals of $1/H(z)$, and thereby to obtain model-independent luminosity distances $d_L(z)$ and distance moduli $\mu_{\rm GP}(z)$.

\subsubsection*{Gaussian Process prior, derivatives, and integration}\label{gp}

A Gaussian Process (GP) defines a distribution over functions such that the values at any set of input points follow a joint Gaussian distribution:
\begin{equation}
f(z) \sim \mathcal{GP}\!\big(\mu(z), K(z,z')\big),
\end{equation}
with mean $\mu(z)$ and covariance kernel $K(z,z')$. We set $\mu(z)=0$ so that the reconstruction is entirely data-driven.  For the covariance, we adopt the squared exponential kernel,  
\begin{equation}
K_{\rm SE}(z,z') = \sigma_f^2 \exp\!\left[-\frac{(z-z')^2}{2\ell^2}\right],
\end{equation}
where $\sigma_f$ and $\ell$ describe the amplitude and correlation length. These hyperparameters are determined from the data by maximizing the GP marginal likelihood:
\begin{equation}
\ln \mathcal{L} = -\tfrac{1}{2}(y-\mu)^{T}[K(Z,Z)+C]^{-1}(y-\mu)
- \tfrac{1}{2}\ln|K(Z,Z)+C| - \tfrac{N}{2}\ln(2\pi),
\end{equation}
with $Z=\{z_i\}$ the observed redshifts, $y$ the $H(z)$ measurements, and $C$ their covariance.  

An advantage of GP regression is that differentiation preserves Gaussianity, so the derivative of a GP-distributed function is also GP-distributed:  
\begin{equation}
\frac{d}{dz}f(z) \;\sim\; \mathcal{GP}\!\left(\mu'(z), 
\frac{\partial^2 K(z,z')}{\partial z \,\partial z'}\right).
\end{equation}
This allows us to obtain both the reconstructed expansion history and its slope in a statistically consistent manner.

This property allows us to compute the derivative $d\Delta M_B/dz$ directly, and to propagate uncertainties into derived quantities such as $1/H(z)$ without relying on noisy finite-difference schemes. In practice, this gives access not only to the reconstructed expansion history but also to its slope, which can reveal departures from smooth evolution.  

The GP reconstruction can also be integrated to obtain comoving distances, since
\begin{equation}
r(z) = \int_0^z \frac{dz'}{H(z')}.
\end{equation}
To propagate uncertainties, we generate Monte Carlo realizations of $H(z)$ from the GP posterior and evaluate them on a Chebyshev grid. The use of Chebyshev nodes reduces interpolation errors and suppresses Runge instabilities, thereby improving numerical stability and quadrature accuracy \citep{karjanto2020, Cap_2018}. This choice ensures that the reconstructed $H(z)$ is well behaved across the redshift range and provides a reliable basis for integration. For each realization, we compute $r(z)$ with the cumulative trapezoidal rule, which is well suited to smooth functions; consistency was verified against Simpson’s 3/8 rule, with negligible differences found.

The luminosity distance and distance modulus follow as
\begin{equation}
d_L(z) = (1+z)\,r(z), \qquad 
\mu_{\rm GP}(z) = 5\log_{10}\!\left(\frac{d_L(z)}{\mathrm{Mpc}}\right)+25.
\end{equation}
Repeating the procedure over many realizations yields the distribution of $\mu_{\rm GP}(z)$, from which we take the mean and standard deviation as the model-independent prediction and its uncertainty.  

In this way GP framework offers three advantages: (i) a smooth, non-parametric reconstruction of $H(z)$ directly from data; (ii) analytic access to derivatives, allowing us to test for redshift-dependent trends; and (iii) consistent propagation of uncertainties into integrated quantities such as $d_L(z)$ and $\mu_{\rm GP}(z)$. These features make it a natural tool for testing the standard candle assumption of Type~Ia supernovae.

\section{Data}\label{Data}

Our analysis is built on three complementary datasets: (i) direct measurements of the Hubble parameter $H(z)$ from cosmic chronometers, which we use to reconstruct a model-independent baseline $\mu_{\rm GP}(z)$; (ii) the Pantheon+ compilation of Type Ia supernovae, which provides the primary observed distance modulus $\mu_{\rm obs}(z)$; and (iii) the Dark Energy Survey 5-Year (DES 5YR) spectroscopic supernova sample, which serves as an independent, homogeneous validation dataset of the result obtained from Pantheon+ compilation. Taken together, these three datasets allow us to test for possible redshift evolution in the intrinsic luminosity of Type Ia supernovae in a way that minimizes cosmological assumptions.

\begin{table}[h]
\centering
\small
\begin{tabular}{lccp{6.5cm}}
\hline
\textbf{Dataset} & \textbf{Size} & \textbf{Redshift Range} & \textbf{Role in Analysis} \\
\hline
Cosmic chronometers ($H(z)$) & 32 points & $0.07 \leq z \leq 1.965$ & Model-independent baseline $\mu_{\rm GP}(z)$ \\
Pantheon+ SNe Ia & 1701 SNe Ia & $0.001 \geq z \geq 2.261$ &  Primary $\mu_{\rm obs}(z)$ anchor  \\
DES 5YR SNe Ia & 435 SNe Ia & $0.02 \leq z \leq 0.72$ & Independent cross-check of Pantheon+ trends \\
\hline
\end{tabular}
\caption{Summary of datasets used in this analysis. The three datasets play complementary roles in constructing and validating the residual diagnostic $\Delta M_B(z)$.}
\label{tab:data_summary}
\end{table}

\subsection*{Cosmic Chronometer $H(z)$ Measurements}

We adopt the most updated observational Hubble data (OHD) catalog, composed of 32 measurements of the Hubble parameter $H(z)$ in the redshift range $0.07 \leq z \leq 1.965$. These data, while affected by relatively large statistical uncertainties, are derived in a completely model-independent manner using the \emph{cosmic chronometer} technique \citep{jimenez_2002, Moresco_2012, Moresco_2016}. 

Cosmic chronometers are passively evolving early-type galaxies that formed the bulk of their stars at high redshift and have since evolved without significant star formation. The differential age method exploits the fact that the relative age difference $\Delta t$ between two such galaxy populations, observed at slightly different redshifts $\Delta z$, directly probes the expansion rate of the Universe. Specifically, one obtains

\begin{equation}
H(z) = -\frac{1}{1+z}\frac{\Delta z}{\Delta t},
\end{equation}
which relates the cosmic expansion rate to observable galaxy chronometry without invoking any cosmological model. 

The robustness of this approach relies on accurate stellar population synthesis (SPS) modeling to determine galaxy ages, and systematics associated with metallicity, star-formation histories, and SPS libraries remain active areas of refinement. Nevertheless, cosmic chronometers provide one of the very few direct and cosmology-independent measurements of $H(z)$ available to date. For complete compilations and methodological details, see Refs.~\citep{Moresco_2016,Jimenez_2023}.

\subsection*{Pantheon+ Supernova Sample}

The Pantheon+ compilation \citep{scolnic2022} represents the most extensive and homogeneous collection of Type~Ia supernovae currently available, comprising 1701 spectroscopically confirmed events across the redshift interval $0.01 \lesssim z \lesssim 2.3$. It unifies observations from Pan-STARRS1, SDSS, SNLS, HST, and multiple low-redshift programs into a self-consistent photometric calibration, thereby establishing the benchmark dataset for cosmological analyses.  

Each supernova light curve is fitted with the SALT2 model \citep{guy_2007}, which yields the rest-frame $B$-band peak magnitude $m_B$, the stretch parameter $X_1$, and the color $C$. Standardization follows the Tripp relation,
\begin{equation}
m_B^{\rm corr} = m_B + \alpha X_1 - \beta C + \Delta_M,
\end{equation}
where $\alpha$ and $\beta$ quantify empirical correlations of luminosity with stretch and color, while $\Delta_M$ accounts for the host-galaxy mass step. The standardized distance modulus is then given by
\begin{equation}
\mu_{\rm obs}(z) = m_B^{\rm corr} - M_B,
\end{equation}
with $M_B$ denoting the fiducial absolute magnitude. Importantly, $M_B$ sets only the overall zero-point of the Hubble diagram: any shift in $M_B$ translates to a uniform vertical offset without altering the relative redshift dependence of $\mu_{\rm obs}(z)$. Thus, while $M_B$ remains degenerate with $H_0$, it does not impact diagnostics of luminosity evolution, which depend on redshift-dependent residuals $\Delta M_B(z) = \mu_{\rm obs}(z) - \mu_{\rm GP}(z)$.  

The Pantheon+ release further provides a full covariance matrix that incorporates both statistical and systematic uncertainties, including calibration errors, Malmquist bias, host-galaxy effects, and intrinsic scatter. This comprehensive error budget underpins its role as the primary reference for testing redshift evolution in Type~Ia supernova luminosities.

\subsection*{DES 5-Year Spectroscopic Supernova Sample}

To complement Pantheon+, we employ the Dark Energy Survey 5-Year (DES 5YR) spectroscopic supernova sample \citep{DES_2024,Camilleri_2024}, consisting of 435 spectroscopically confirmed SNe~Ia in the redshift range $0.02 \leq z \leq 0.72$. Unlike Pantheon+, which synthesizes multiple heterogeneous surveys, DES is a single, homogeneous program conducted with the Dark Energy Camera (DECam) on the Blanco 4m telescope. Its uniform observing strategy, calibration pipeline, and well-defined selection function minimize cross-survey systematics and yield an internally consistent Hubble diagram.  

As in Pantheon+, light curves are fitted with SALT2, and standardized using the Tripp relation (Eq.~1), with $\alpha$, $\beta$, and the host mass step marginalized over. The observed distance modulus is defined identically (Eq.~2), with $M_B$ serving only as a global zero-point.  

In our analysis, the DES 5YR dataset serves as an independent validation of the Pantheon+ results in the overlapping redshift regime. Because DES is both smaller in size and narrower in redshift coverage, it does not replace Pantheon+ as the primary anchor; rather, it provides a critical cross-check \citep{Vincenzi_2025}. If the same $\Delta M_B(z)$ structures inferred from Pantheon+ are reproduced in DES, the case for genuine luminosity evolution is strengthened. Conversely, any systematic discrepancy between the two samples would point to survey-dependent calibration issues or selection effects, thereby helping us disentangle astrophysical signals from observational systematics.


\section{Results}\label{Result}

\begin{figure}[h]
    \centering
    \includegraphics[width=0.32\linewidth]{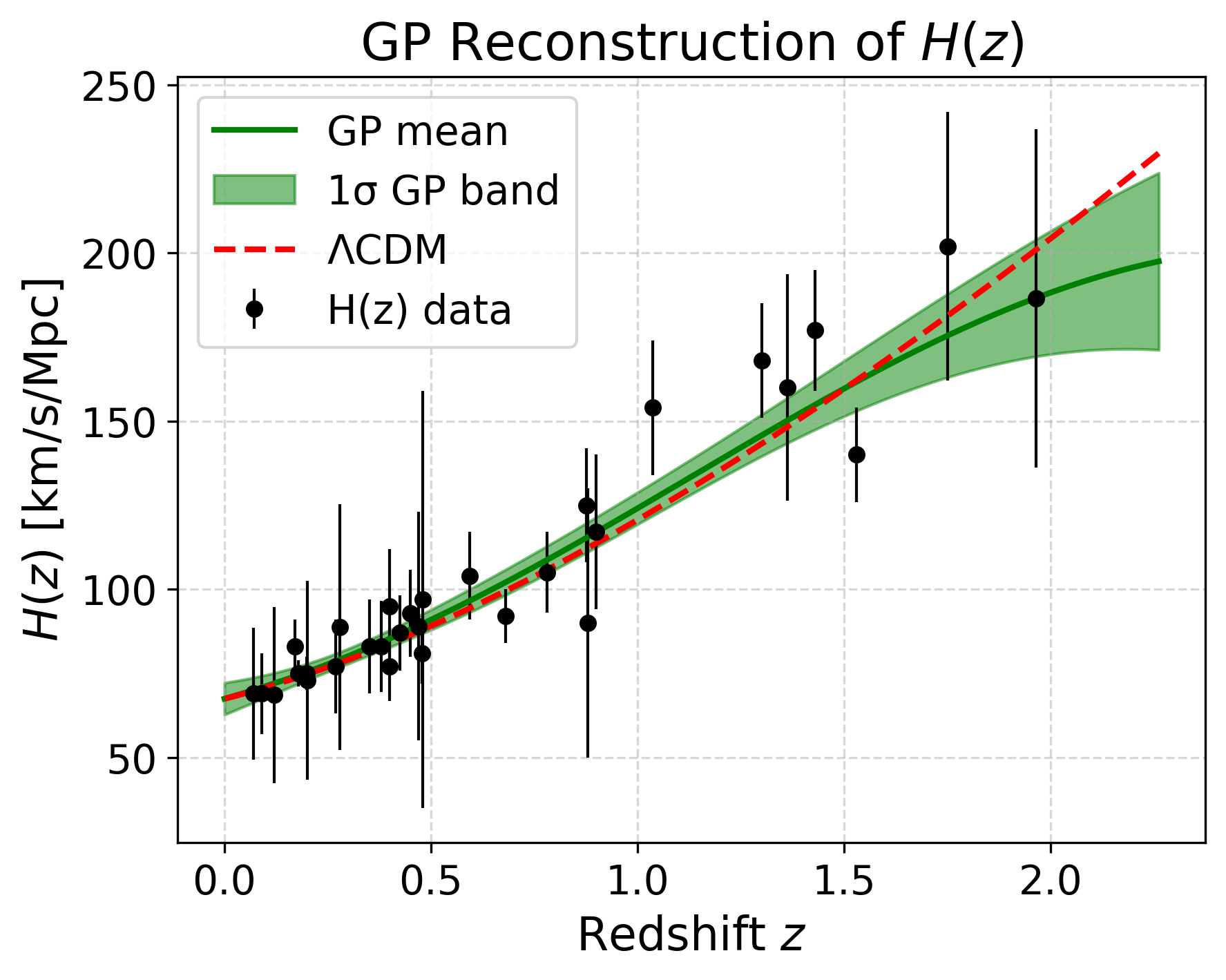}
    \includegraphics[width=0.32\linewidth]{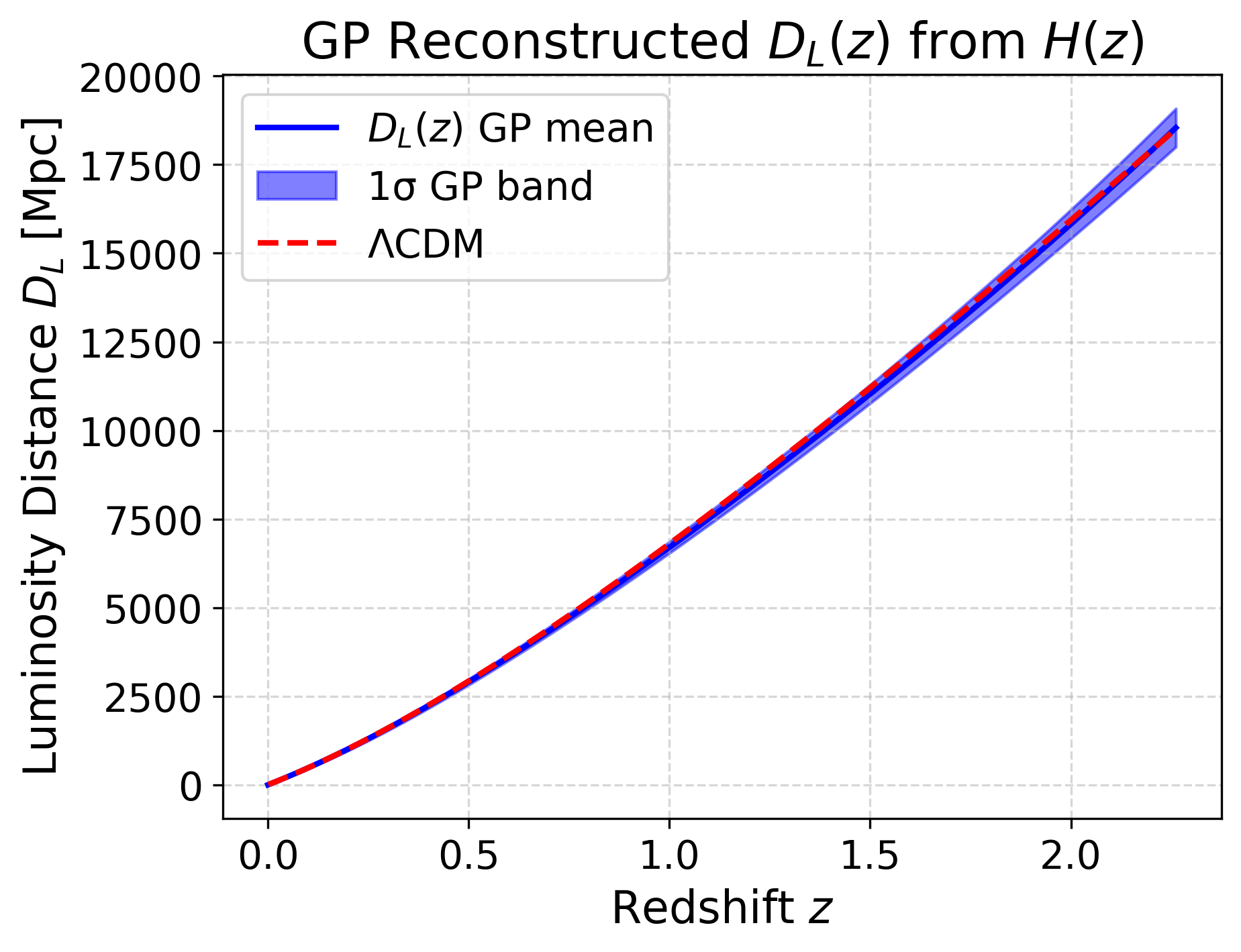}
    \includegraphics[width=0.32\linewidth]{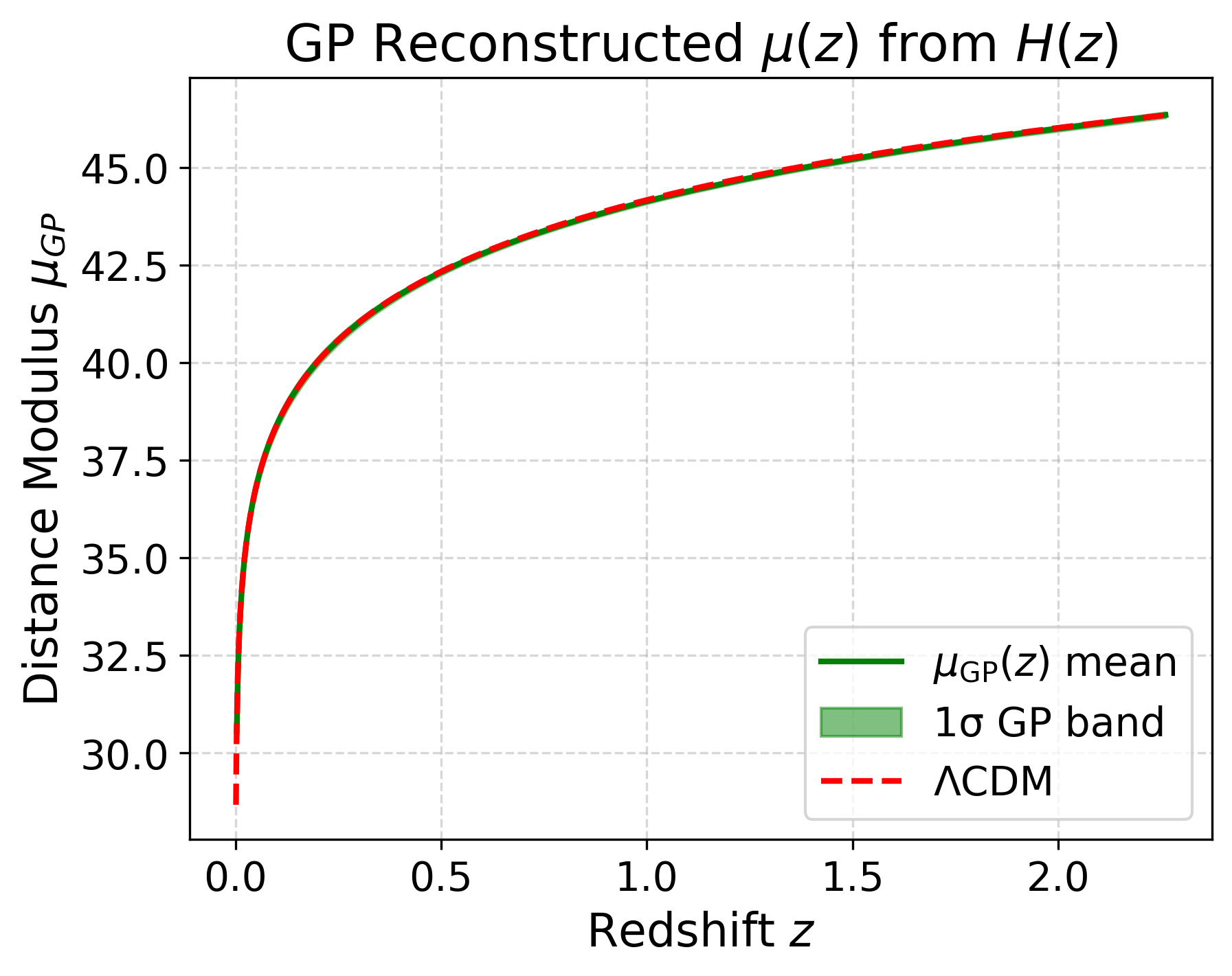}
    \caption{Gaussian Process reconstruction of $H(z)$ from cosmic chronometer data (left), the derived luminosity distance $D_L(z)$ (middle), and the corresponding distance modulus $\mu(z)$ (right), shown alongside the $\Lambda$CDM prediction for comparison.}
    \label{hztomugp}
\end{figure}

\begin{figure}[h]
    \centering
    \includegraphics[width=0.49\linewidth]{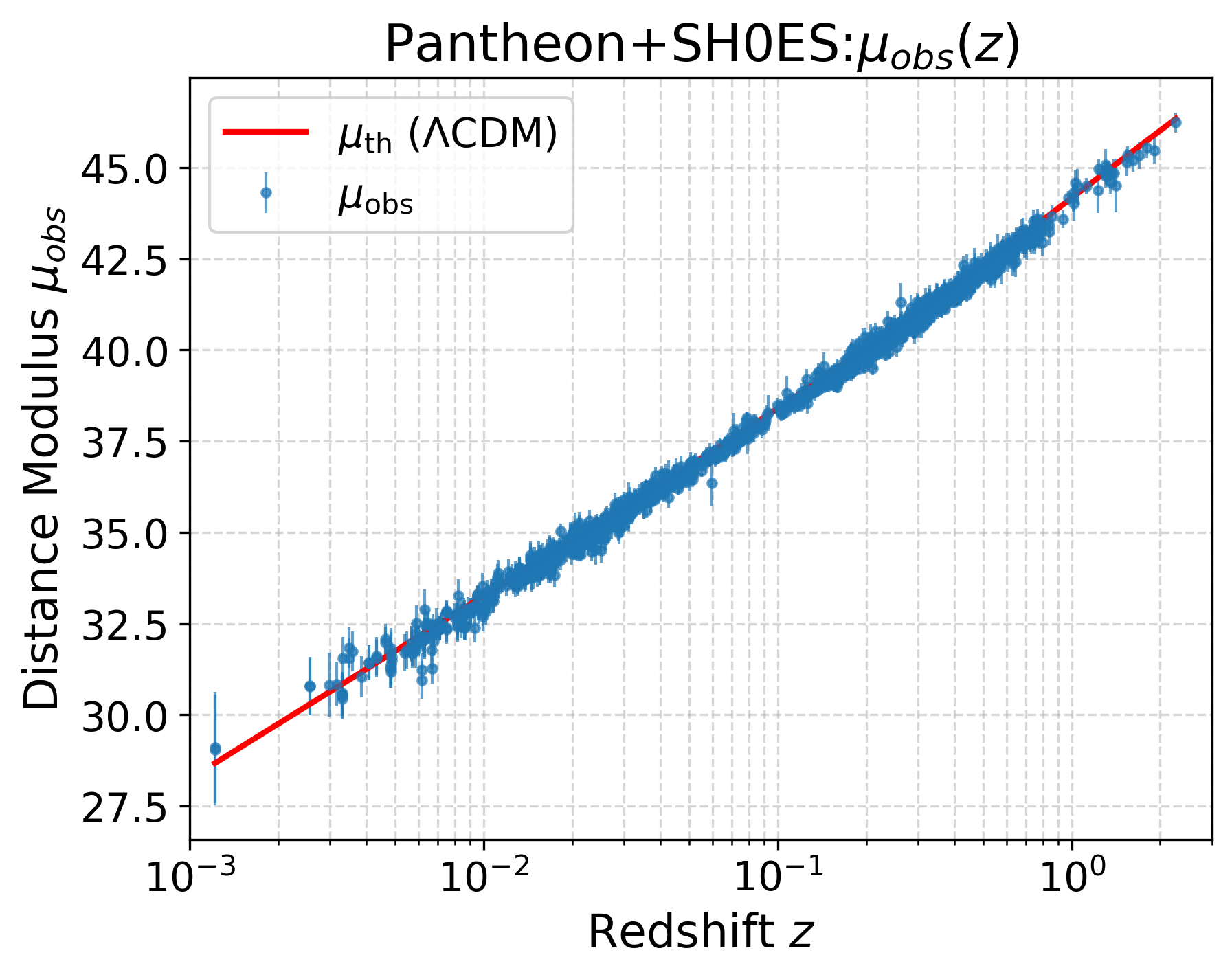}
    \includegraphics[width=0.49\linewidth]{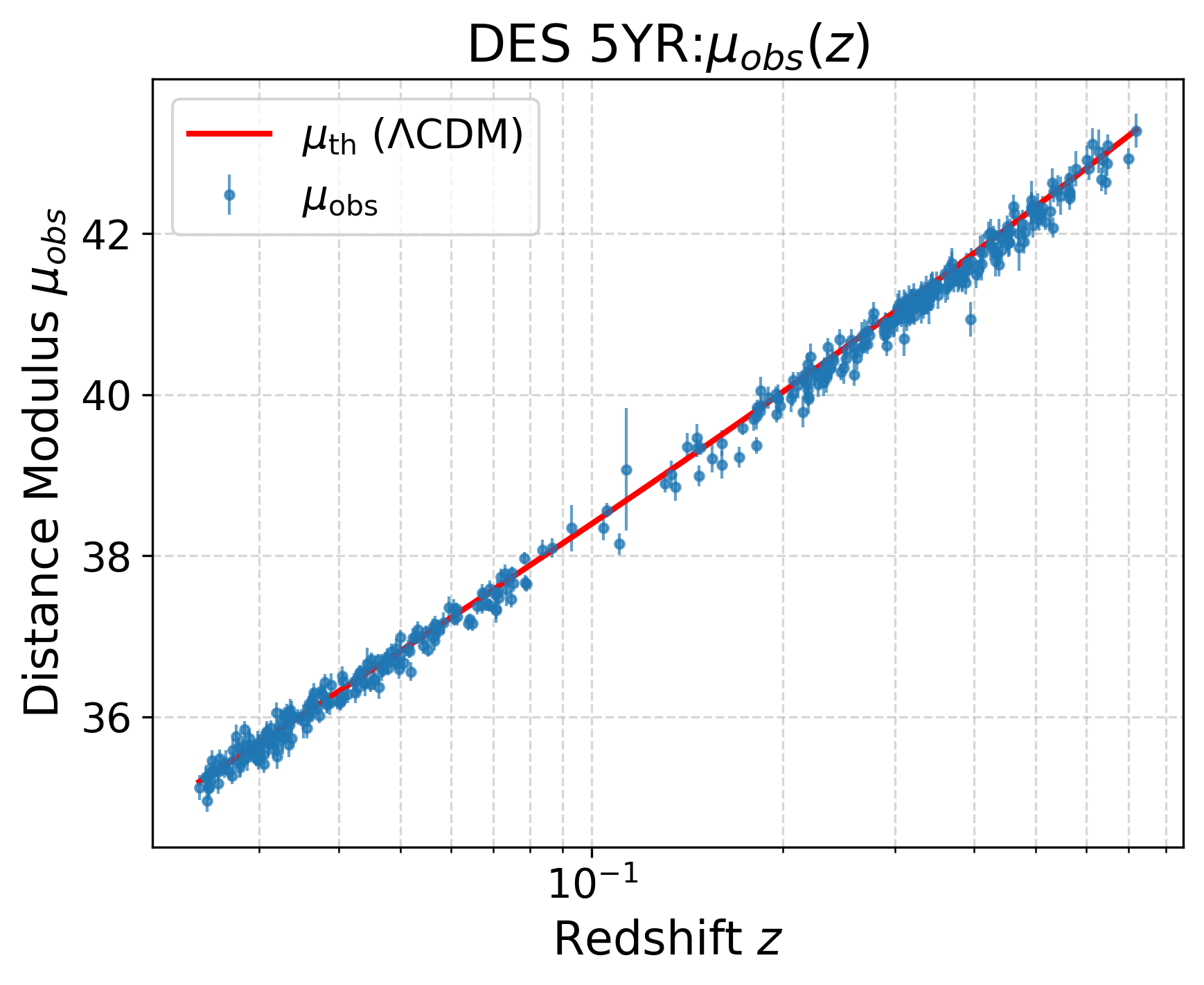}
    \caption{Observed distance modulus $\mu_{\rm obs}(z)$ for Pantheon+ (left) and DES 5YR (right), compared with the fiducial $\Lambda$CDM prediction $\mu_{\rm th}(z)$.}
    \label{mu_obs}
\end{figure}

\begin{figure}[h]
    \centering
    \includegraphics[width=0.49\linewidth]{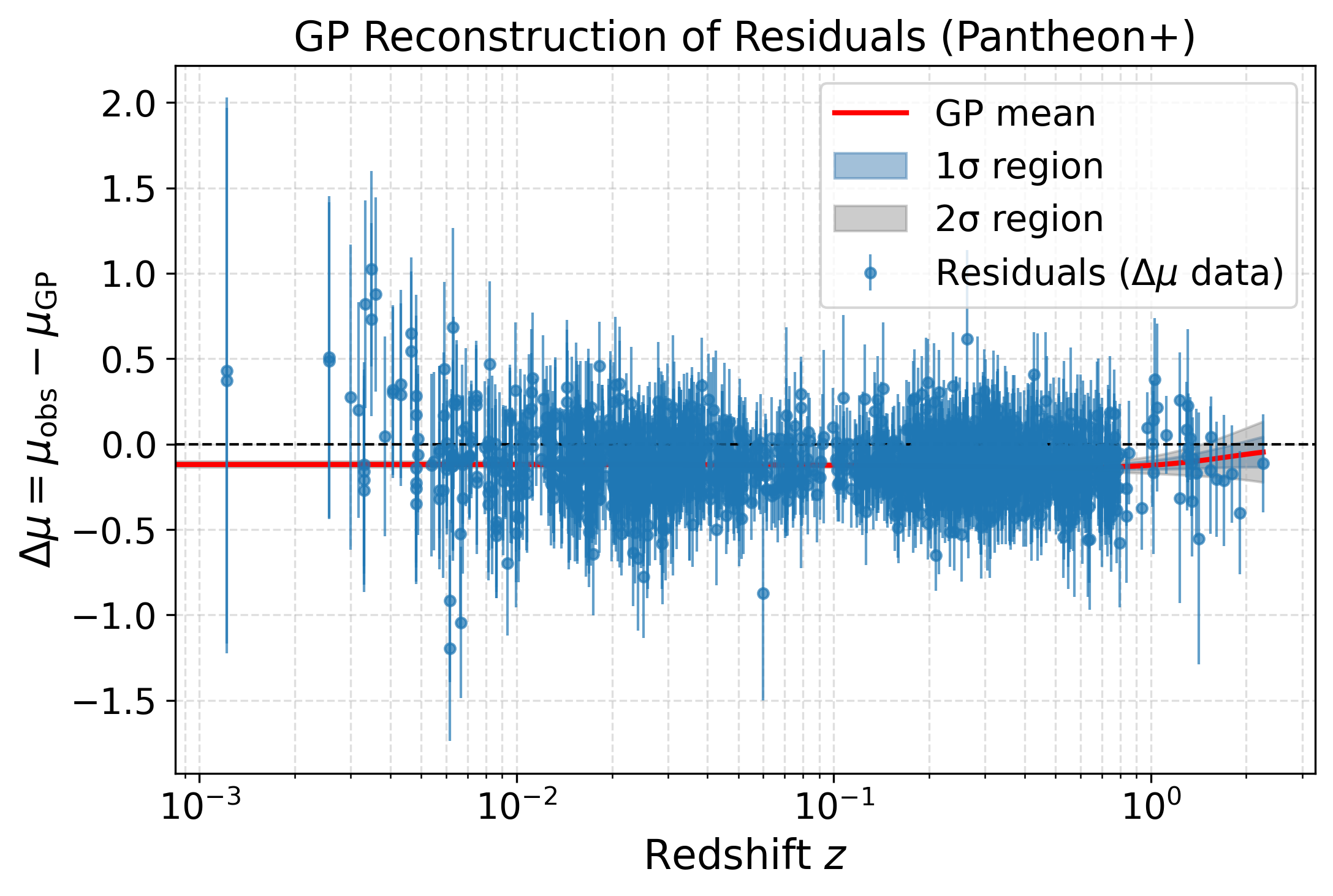}
    \includegraphics[width=0.49\linewidth]{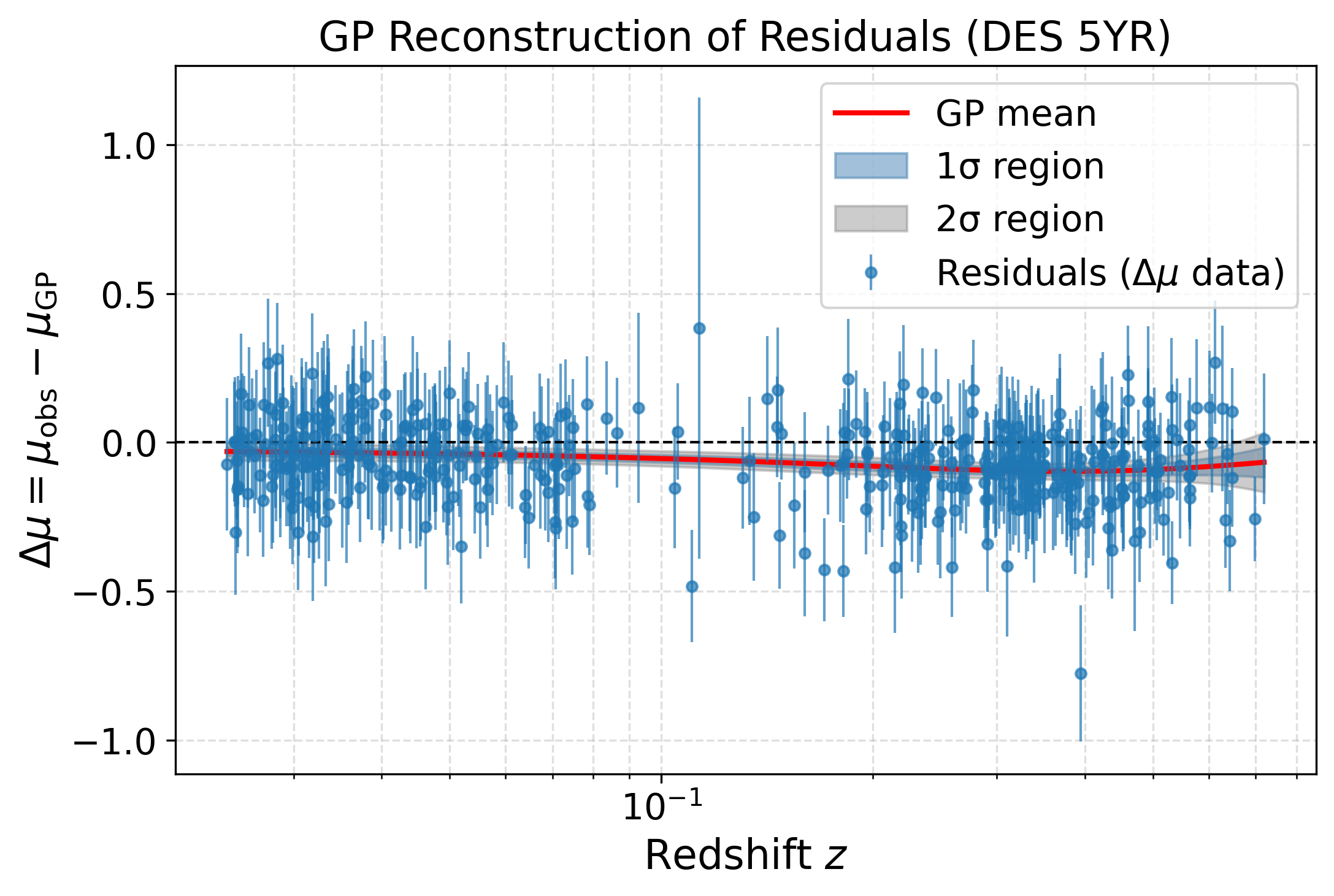}
    \caption{GP reconstruction of supernova residuals $\Delta\mu = \mu_{\rm obs} - \mu_{\rm GP}$ for Pantheon+ (left) and DES 5YR (right), compared with the GP mean and confidence regions.}
    \label{delta_mu}
\end{figure}

\begin{figure}
    \centering
    \includegraphics[width=0.49\linewidth]{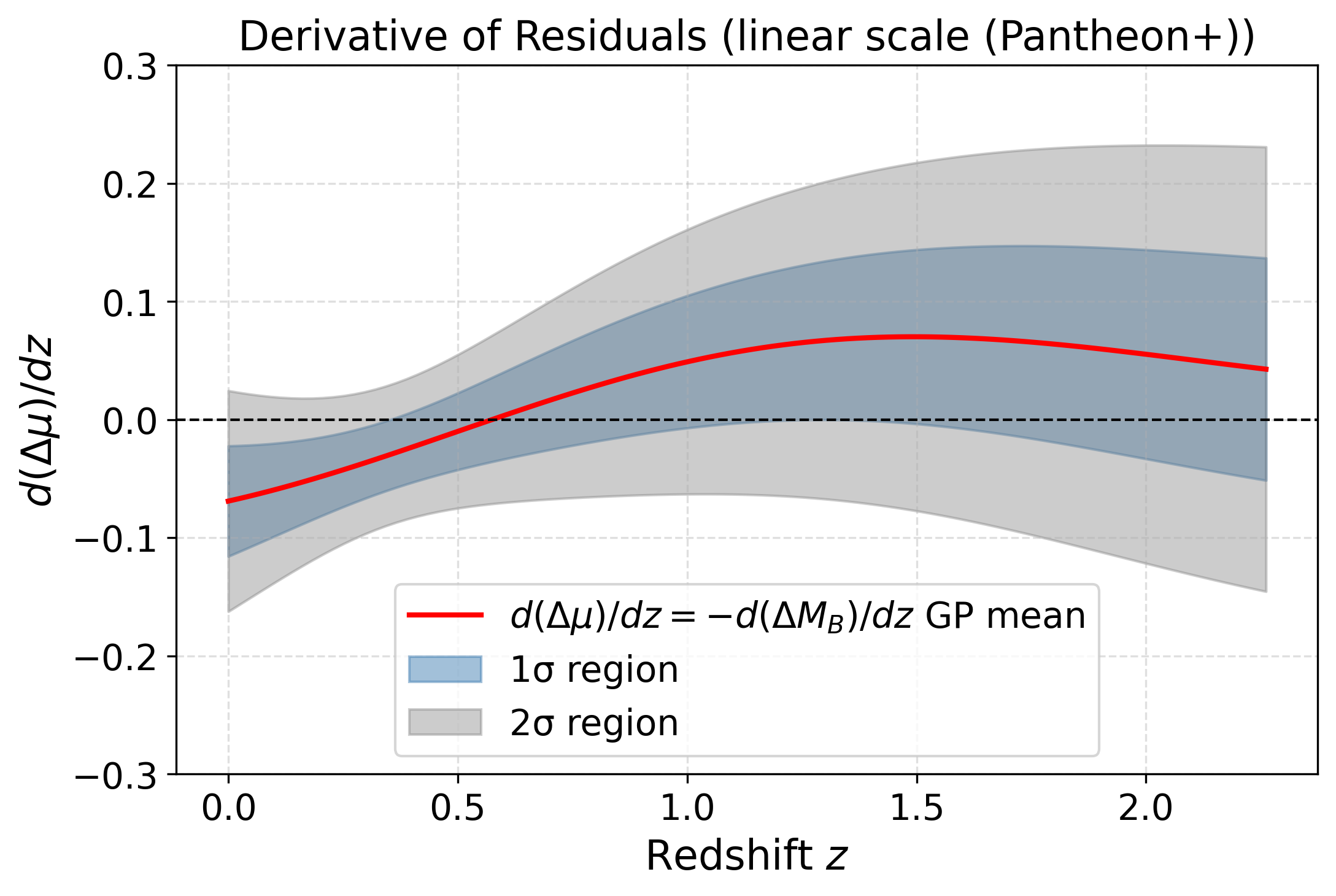}
    \includegraphics[width=0.49\linewidth]{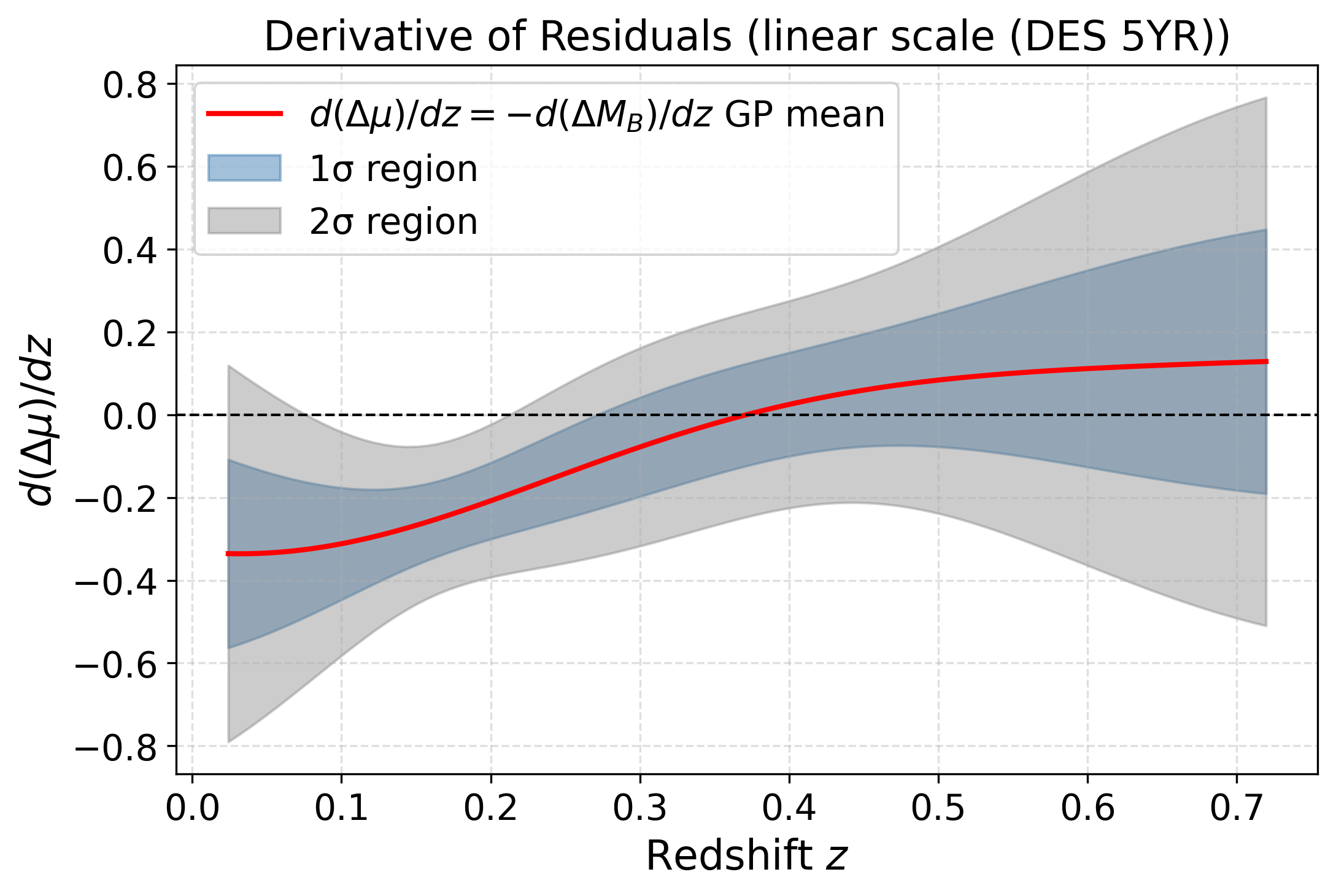}
    \caption{Derivative of residuals $d(\Delta\mu)/dz$ reconstructed with Gaussian Processes for Pantheon+ (left) and DES 5YR (right), shown on a linear redshift scale with $1\sigma$ and $2\sigma$ confidence regions.}
    \label{d_dz}
\end{figure}

The results of our analysis are summarized in Figures~\ref{hztomugp}-\ref{d_dz}, where we present a sequence of reconstructions aimed at testing the standard candle property of Type~Ia supernovae in a model- independent framework. We begin with the reconstruction of the cosmic expansion history from 32 cosmic chronometer measurements of the Hubble parameter, $H(z)$. Using Gaussian Processes (see Section~\ref{gp}), we obtain a smooth, non-parametric interpolation of the data, with the central curve representing the mean reconstruction and the shaded regions denoting the one and two sigma confidence intervals. This reconstruction is displayed in the left panel of Figure~\ref{hztomugp}.   From this reconstructed expansion history, we derive the luminosity distance by numerical integration. For this step we assume a spatially flat universe, consistent with current cosmological constraints. The luminosity distance is then converted into the distance modulus, which provides a direct comparison with supernova observations. 

To capture both statistical noise and correlated systematics present in the chronometer data, we employ the full covariance matrix and generate 2000 Monte Carlo realizations of the $H(z)$ measurements. Each realization undergoes the full reconstruction and integration pipeline, yielding an ensemble of distance modulus curves. The corresponding distance modulus $\mu_{\mathrm{GP}}(z)$ is  presented in the right panel of Figure~\ref{hztomugp}, allowing a direct comparison with Type~Ia supernova observations.   The spread across these realizations defines our error estimates, which are therefore comprehensive and statistically sound. This agreement between the reconstructed plot and $\Lambda$CDM model validates our model-independent Gaussian Process reconstruction as a reliable baseline. It also establishes a cosmology-independent reference against which the supernova distance moduli can be compared, enabling a direct test for possible redshift evolution in the intrinsic luminosity of Type~Ia supernovae.

In Figure~\ref{mu_obs} we present the observed Hubble diagrams for the Pantheon+ and DES-SN5YR samples, where the distance moduli $\mu_{\mathrm{obs}}(z)$ are plotted against redshift. The Pantheon+ compilation, comprising 1701 supernovae spanning $0.001 \leq z \leq 2.3$, yields a densely sampled and statistically powerful diagram across a broad redshift range. The DES 5YR sample, while smaller and limited to $z < 0.72$, provides an independent and internally homogeneous dataset that is valuable for cross-validation. Both datasets broadly trace the $\Lambda$CDM prediction, confirming the robustness of the standard model in describing the mean expansion history. However, such agreement only tests mean distances; it cannot exclude subtle redshift-dependent drifts in intrinsic luminosities. This motivates residual-based diagnostics.

Further, we compute residuals relative to a cosmology-independent baseline provided by Gaussian Process (GP) reconstructions of $\mu(z)$ derived from cosmic chronometer data (Section~\ref{gp}). The residuals are defined as  
\[
\Delta \mu(z) = \mu_{\mathrm{obs}}(z) - \mu_{\mathrm{GP}}(z).
\]  
Figure~\ref{delta_mu} shows the resulting distributions of $\Delta \mu(z)$ for Pantheon+ (left) and DES 5YR (right). Pantheon+ residuals cluster tightly around zero, with the GP mean lying well within the $1\sigma$ band across most of the redshift range. DES residuals show larger scatter due to smaller sample size but remain broadly consistent with Pantheon+. Both datasets, however, reveal localized departures, particularly near $z\sim 1$ (Pantheon+) and $z\sim 0.3$--0.5 (DES). These features, although small, are statistically significant enough to warrant further scrutiny.

A more incisive diagnostic comes from analyzing the  derivative of the residuals, $d(\Delta \mu)/dz$, shown in Figure~\ref{d_dz}, provides a direct, data-driven probe of possible redshift evolution in the standardized luminosities of Type~Ia supernovae. In this framework: (i) statistical consistency of the mean trend with zero supports the constant-$M_B$ assumption, (ii) systematic monotonic drifts would point toward genuine luminosity evolution, and (iii) localized departures could indicate subtler systematics tied to progenitor channels, host-galaxy environments, or intergalactic opacity effects.  For Pantheon+, the derivative remains broadly consistent with zero but exhibits a mild positive excursion around $z \sim 1$, suggesting that SNe~Ia may be marginally brighter at higher redshifts than expected for a strictly constant absolute magnitude. DES shows a weaker but qualitatively similar slope between $z \sim 0.3$--0.5. The consistency of these features across independent surveys supports their credibility. Importantly, the derivative-based diagnostic enhances sensitivity to redshift-dependent structure that could otherwise be masked in cumulative residuals.

Overall, the results establish that while SNe~Ia are standardizable to high precision, subtle redshift-dependent features are present in both Pantheon+ and DES. The qualitative agreement between the two datasets in the intermediate-redshift regime strengthens the case that the observed deviations are real features, not statistical fluctuations or calibration artifacts. These features form the basis of the discussion below.


\section{Discussion}\label{Discussion}

The results point to a two-fold conclusion. First, Type~Ia supernovae continue to behave as remarkably reliable standardizable candles, with deviations constrained to the level of only a few hundredths of a magnitude [see also \citep{Brout_2022}]. Second, within this overall consistency, both Pantheon+ and DES-SN5YR exhibit identical localized departures from strict constancy. 

An interesting feature of our diagnostic is the behaviour of the slope $d(\Delta\mu)/dz$ across redshift. At low redshift, both Pantheon+ and DES 5YR show negative values of $d(\Delta\mu)/dz$. Since $d(\Delta\mu)/dz = -\,dM_B/dz$, this corresponds to $dM_B/dz > 0$, implying that the absolute magnitude increases with redshift and supernovae become progressively fainter relative to their assumed constant value. Such a trend is consistent with theoretical expectations that lower metallicity or younger progenitors at earlier epochs may synthesize less $^{56}$Ni, reducing their peak luminosity \citep{Timmes_2003,How_2009,Kim_2025}. This interpretation is also supported by empirical evidence that older progenitor populations yield brighter standardized SNe~Ia: \citet{Lee_2021,Krueger_2010} find strong progenitor-age dependence in the width-luminosity and color-luminosity relations, while \citet{Wiseman_2023} show that events in massive, passive galaxies are systematically brighter, and \citet{Wang_2023} confirm a similar though weaker correlation. 

Beyond a certain redshift, however, the slope changes sign and becomes positive. This corresponds to $dM_B/dz < 0$, i.e.\ a decreasing absolute magnitude with redshift, such that supernovae appear brighter at higher $z$. A natural astrophysical explanation is a transition in the dominant progenitor channel: at intermediate redshifts, younger and more massive systems in actively star-forming environments could produce intrinsically brighter explosions, making the observed SNe appear overluminous compared to the low-$z$ baseline\citep{Sull2010,Kim_2019}. The combination of a faintening trend at low redshift and a brightening trend at higher redshift therefore points to non-monotonic luminosity evolution, with different physical drivers operating in distinct redshift regimes.

While survey-specific effects such as calibration offsets, K-corrections, or selection biases could mimic such behaviour, the fact that Pantheon+, a heterogeneous compilation, and DES-SN5YR, a homogeneous, well-controlled survey, both display the same qualitative excursions reduces the likelihood that the observed patterns are merely statistical fluctuations. Their consistency across independent datasets strengthens the case that the signal is robust and warrants interpretation in terms of progenitor evolution and its possible cosmological consequences. In particular, the agreement between Pantheon+ and DES at intermediate redshift suggests that at least part of the trend may be astrophysical. At the same time, since both datasets rely on SALT2 with fixed nuisance parameters $(\alpha,\beta,\Delta_M)$, any genuine dependence on redshift or host properties could also appear as evolution in $\Delta M_B(z)$. Our method therefore provides a model-independent means of flagging such departures. Cosmologically, even a modest drift of $\sim0.05$ mag in $\Delta M_B(z)$ introduces a percent-level bias in distance moduli, sufficient to shift inferred values of $H_0$ and $w$\cite{Campbell_2016,Linden_2009}. This underscores the importance of modeling luminosity evolution carefully in future Stage~IV surveys \citep{Palan2025}, where systematic errors must be controlled below statistical precision.

In summary, this work highlights the strength of non-parametric Gaussian Process (GP) methods for precision cosmology. Unlike parametric models, which impose fixed functional forms on possible luminosity evolution, the GP approach allows the data to reveal both gradual trends and localized features, improving sensitivity to subtle effects. By anchoring the baseline in cosmic chronometer measurements, we reduce dependence on cosmological assumptions and provide a robust, independent test of the supernova standard candle paradigm. More broadly, the GP framework serves as a flexible bridge between different cosmological probes, complementing both parametric fits and hierarchical population models. Its successful application to Pantheon+ and DES 5YR demonstrates its value as a general tool, one that can be directly extended to future surveys with larger samples and more precise calibration.

\section*{Acknowledgements}

The author wishes to express sincere gratitude to the Principal of St.~Stephen's College and to the Centre for Theoretical Physics, St.~Stephen's College, for their continuous support and for providing the necessary academic infrastructure. The author also gratefully acknowledges Prof.~Shobhit Mahajan and Prof.~Deepak Jain for their valuable guidance.

\bibliographystyle{unsrtnat}
\bibliography{ref}

\end{document}